\documentclass[a4paper,11pt]{article}
\usepackage{pos}
\usepackage{epstopdf}
\usepackage{overpic}
\usepackage{lineno}

\graphicspath{./}

\def\Dz {D^{0}}

\def\pip {\pi^{+}}
\def\pim {\pi^{-}}
\def\piz {\pi^{0}}

\def\Ks {K_S^0}
\def\Kp {K^{+}}
\def\Km {K^{-}}

\title{Recent charm results from Belle}

\author*{Longke Li}

\affiliation{
{\bf On behalf of the Belle Collaboration}\\
University of Cincinnati,\\
  Cincinnati, Ohio 45221, U.S.}

\emailAdd{lilk@ucmail.uc.edu}

\abstract{Recent charm results from Belle experiment are presented in this proceedings, including 
(1) measurement of mixing parameter $y_{CP}=(0.96\pm0.91\pm 0.62^{+0.17}_{-0.00})\%$ in $CP$-odd decay for the first time, 
(2) the first Dalitz-plot analysis of $D^0\to K^-\pi^+\eta$, 
(3) measurement of branching fractions of $\Lambda_c^+\to\eta\Lambda^0\pi^+$ and $\eta\Sigma^0\pi^+$ and intermediate processes $\Lambda_c^+\to[\Lambda(1670)\to\eta\Lambda^0]\pi^+$ and $\Lambda_c^+\to\eta\Sigma(1385)^+$ relative to $\Lambda_c^+\to p\Km\pip$: $0.293\pm0.003\pm0.014$, $0.120\pm0.006\pm0.006$, $(5.54\pm0.29\pm0.73)\times10^{-2}$, and $0.192\pm 0.006\pm 0.016$, respectively, 
and (4) first determination of the spin and parity of a charmed-strange baryon $\Xi_c(2970)^+$ which is consistent with the HQSS prediction for $J^P(s_l)=1/2^+(0)$.}

\FullConference{%
  40th International Conference on High Energy physics - ICHEP2020\\
  July 28 - August 6, 2020\\
  Prague, Czech Republic (virtual meeting)
}

\begin{document}
\maketitle

\section{Introduction to Belle at KEKB}
KEKB~\cite{bib:KEKB} is an asymmetric-energy $e^+e^-$ collider operating at and near $\Upsilon(4S)$ mass peak. As the only detector installed in KEKB, Belle detector has a good performance on momentum and vertex resolution, $K/\pi$ separation etc. A detailed description of the Belle detector can be found elsewhere~\cite{bib:BelleDetector}. It has been ten years since the final full data set ($\sim$1 ${\rm ab^{-1}}$)  was accumulated, however, fruitful results on physics are lasting to be produced. Here we select some recent charm results from Belle to present in this proceedings. 

\section{Charm-mixing parameter $y_{CP}$ in $D^0\to\Ks \omega$}
The mixing parameter $y_{CP}$ is measured in $D^0$ decays to the $CP$-odd final state $\Ks\omega$ for the first time~\cite{bib:ycp_D0ToKsOmega}.
Considering mixing parameters $|x|$ and $|y|\ll1$, the decay-time dependence of $D^0$ to a $CP$ eigenstate is approximately exponential, $d\Gamma/dt \propto e^{- \Gamma(1+\eta_{f} y_{cp})t}$ where $\eta_f=+1$ ($-1$) for $CP$-even (-odd) decays.
Along with the decay rate in flavored eigenstate decays $d\Gamma/dt \propto e^{-\Gamma t}$, the $y_{CP}$ is determined by the decay proper-time value with the formula $y_{CP}=1-\frac{\tau(\Dz\to\Km\pip)}{\tau(\Dz\to\Ks\omega)}$, where $\Dz\to\Km\pip$ is the chosen normalization mode with flavor eigenstate final state. 

\begin{figure}[!htpb]
  \begin{centering}
  \begin{overpic}[width=0.49\textwidth,height=0.48\textwidth]{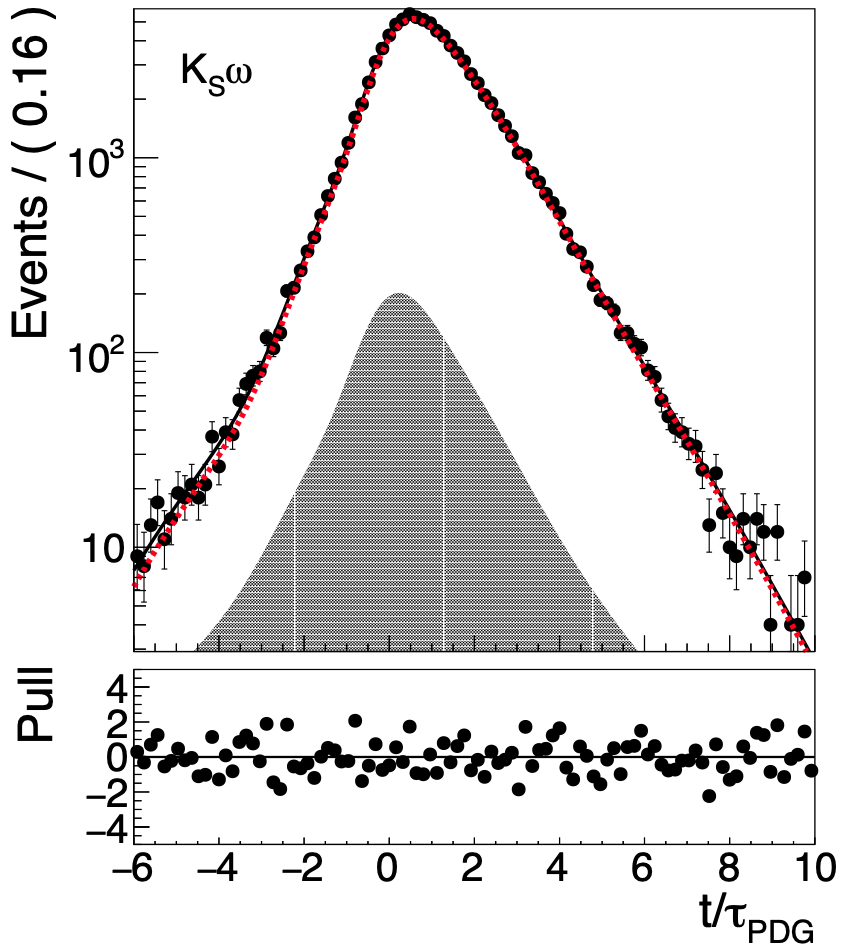}
  \put(20, 64){\large(a)}
  \end{overpic}~~
  \begin{overpic}[width=0.49\textwidth,height=0.48\textwidth]{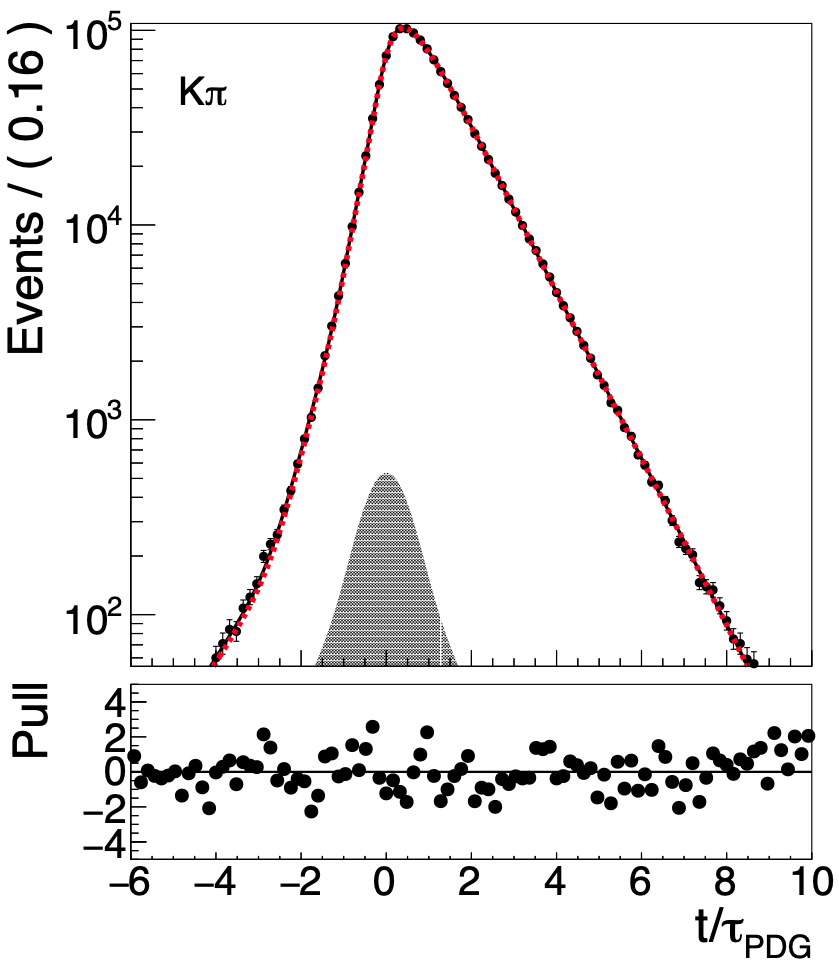}
  \put(20, 64){\large(b)}
  \end{overpic}
  \vskip-5pt
  \caption{\label{fig:ycp} The fit of $D^0$ proper lifetime: (a) $D^0\to\Ks\omega$ and (b) $D^0\to\Km\pip$. The dashed red curves are the signal contribution, and the shaded surfaces beneath are the background estimated from $M-\Delta M$ sidebands.}
  \end{centering}
\end{figure}

 Based on the full Belle data sample of 976 $\rm fb^{-1}$, we obtain 91 thousands of $\Dz\to\Ks\omega$ and 1.4 millions of reference mode $\Dz\to\Km\pip$ in $M-\Delta M$ signal region, where $M$ is the invariant mass of reconstructed $\Dz$ and $\Delta M$ is the mass difference of reconstructed $D^{*+}$ and $\Dz$. 
Using unbinned maximum-likelihood fits for lifetime on these two samples with high purities, the proper decay-time of $\Dz$ is determined as $\tau_{\Ks\omega}=(410.47\pm 3.73)$ fs and $\tau_{K\pi}=(406.53\pm 0.57)$ {\rm fs}, as shown in Fig.~\ref{fig:ycp}. 
Thus, we calculate $y_{CP}=(0.96\pm0.91\pm 0.62^{+0.17}_{-0.00})\%$, where the first uncertainty is statistical, the second is systematic due to event selection and background, and the last is due to possible presence of CP-even decays in the data sample. 
This $y_{CP}$ result is consistent with the world average value. 
In the future, comparing more precise measurements of $y_{CP}$ with that of $y$ may test the SM precisely or reveal new physics effects in the charm system.

\section{Dalitz-plot analysis of $\Dz\to\Km\pip\eta$ decays}
\begin{figure}[!htpb]
  \begin{centering}
  \begin{overpic}[width=0.45\textwidth]{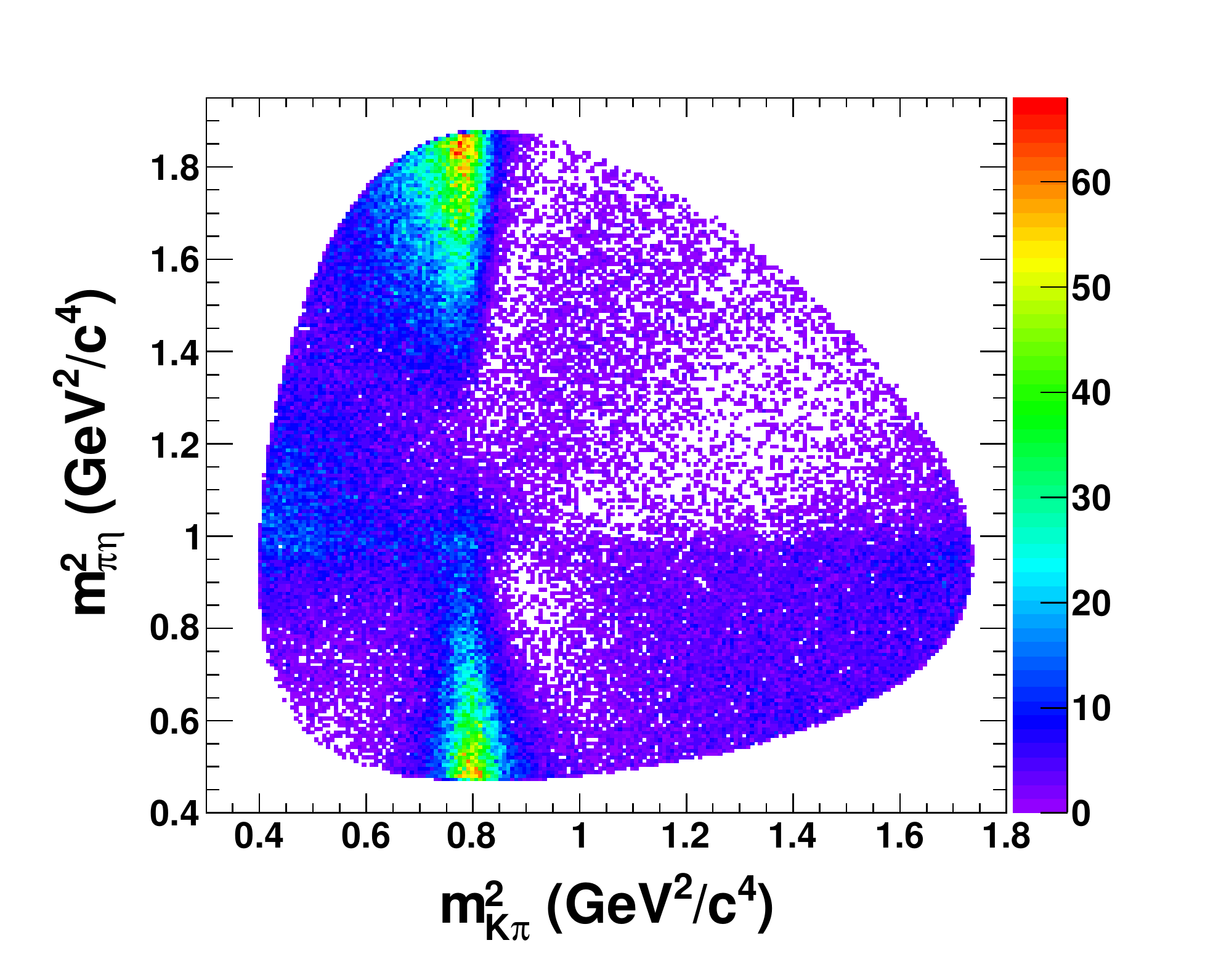}
  \put(21, 65){\large(a)}
  \end{overpic}%
  \begin{overpic}[width=0.42\textwidth]{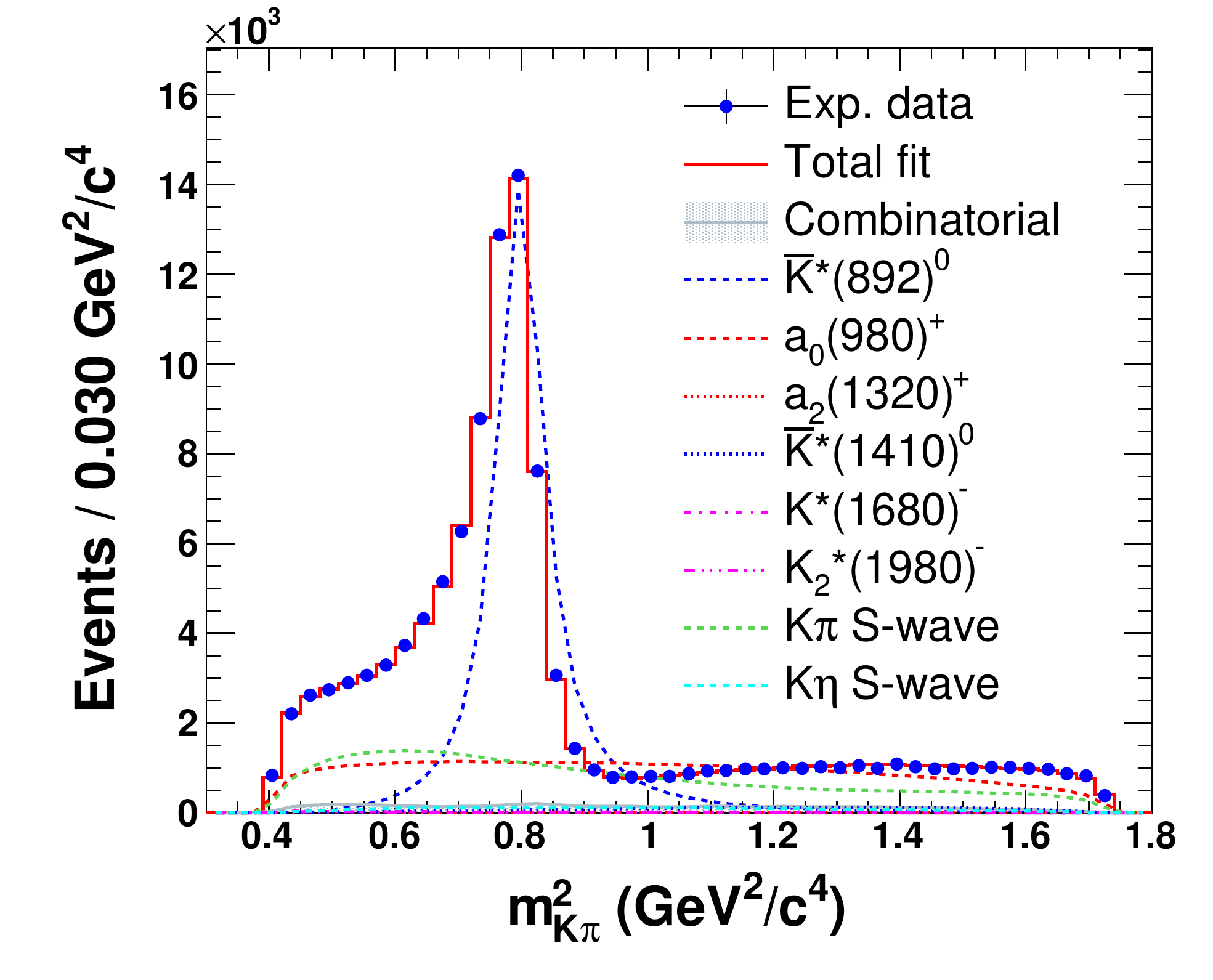}
  \put(22, 68){\large(b)}
  \end{overpic}\\
  \begin{overpic}[width=0.42\textwidth]{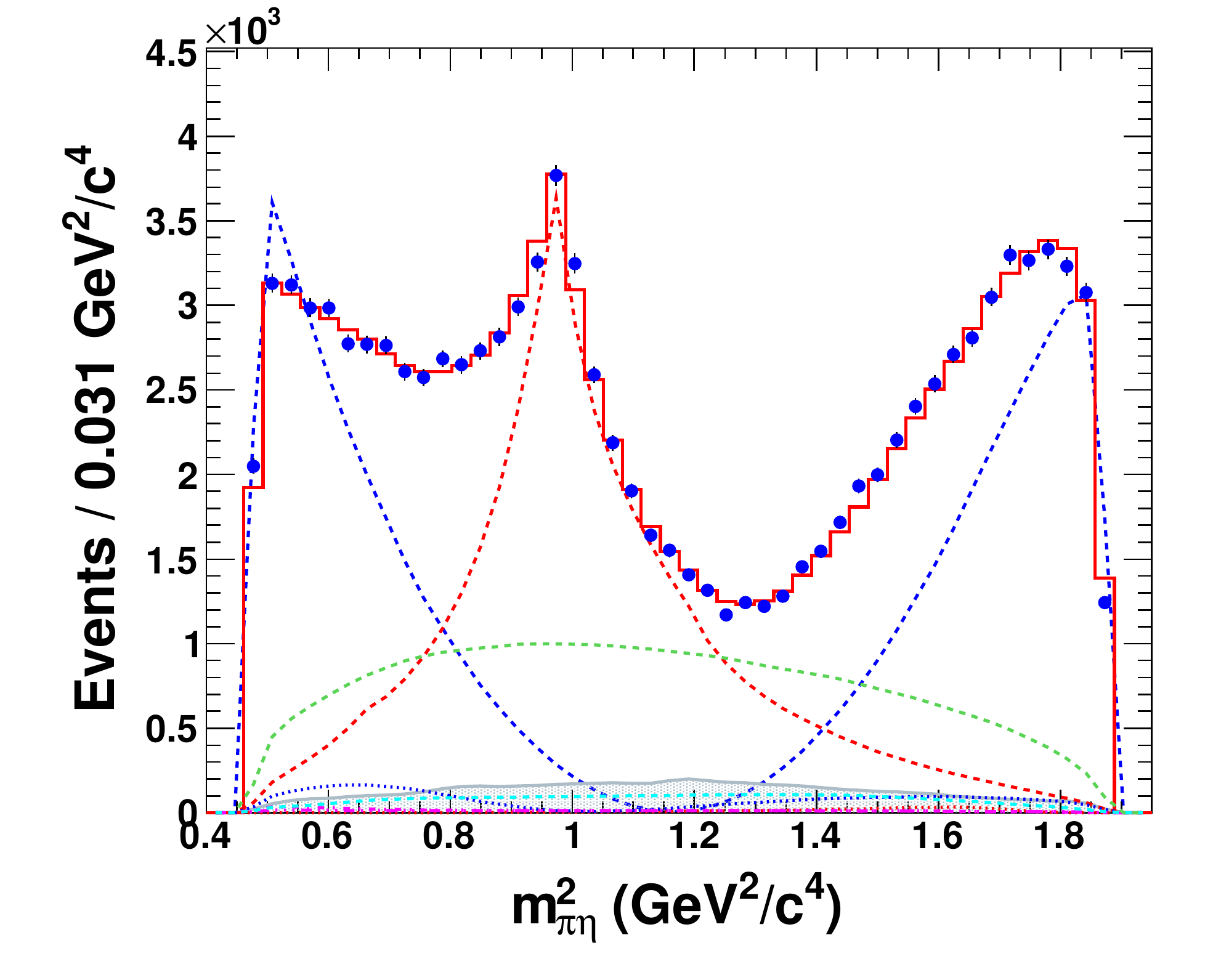}
  \put(22, 68){\large(c)}
  \end{overpic}%
  \begin{overpic}[width=0.42\textwidth]{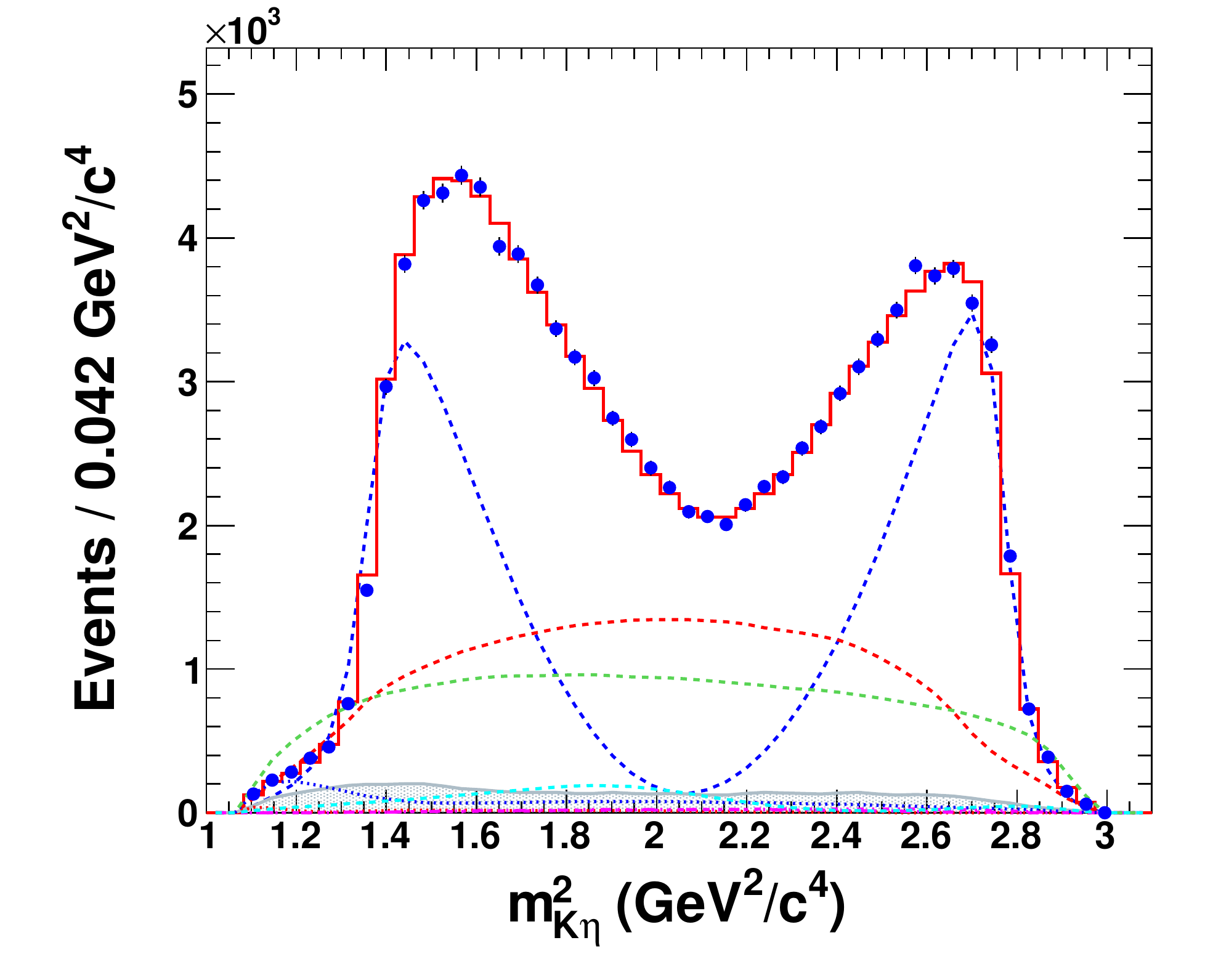}
  \put(22, 68){\large(d)}
  \end{overpic}%
  \vskip-5pt
  \caption{\label{fig:nominal} The Dalitz plot of $\Dz\to\Km\pip\eta$ in (a) $M$-$Q$ signal region $1.85~{\rm GeV}/c^2 < M < 1.88~{\rm GeV}/c^2$ 
and $5.35~{\rm MeV}/c^2 < Q < 6.35~{\rm MeV}/c^2$, and projections on (b) $m_{K\pi}^2$, (c) $m_{\pi\eta}^2$ and (d) $m_{K\eta}^2$. In projections the fitted contributions of individual components are shown, along with contribution of combinatorial background (grey-filled) from sideband region.}
  \end{centering}
\end{figure}

The understanding of hadronic charmed-meson decay is theoretically challenging due to the significant non-perturbative contributions, and input from experimental measurements thus plays an important role. A Dalitz-plot analysis of $D^0\to\Km\pip\eta$ is performed for the first time at Belle based on 953 ${\rm fb^{-1}}$ of data~\cite{bib:PRD102012002}. 
Using a $M$-$Q$ two-dimensional fit where $M$ is the invariant-mass of reconstructed $D^0$ meson, $M=M(\Kp\pim\eta)$, and $Q$ is the released energy of $D^{*+}$ decay, $Q=M(\Km\pip\eta\pi_s)-M-m_{\pi_s}$, a signal yield of $105\,197\pm990$ is obtained in the signal region of $1.85~{\rm GeV}/c^2 < M < 1.88~{\rm GeV}/c^2$ 
and $5.35~{\rm MeV}/c^2 < Q < 6.35~{\rm MeV}/c^2$ with a high purity $(94.6\pm0.9)\%$. 
The Dalitz plot is well described by a combination of the six resonant decay channels $\bar{K}^{*}(892)^0\eta$, $\Km a_0(980)^+$, $\Km a_2(1320)^+$, $\bar{K}^{*}(1410)^0\eta$, $K^{*}(1680)^-\pip$ and  $K_2^{*}(1980)^-\pip$, together with  $K\pi$ and $K\eta$ S-wave components, as shown in Fig.~\ref{fig:nominal}.
The dominant contributions to the decay amplitude arise from $\bar{K}^{*}(892)^{0}$, $a_0(980)^{+}$ 
and the $K\pi$ S-wave component. The $K\eta$ S-wave component, including $K_0^{*}(1430)^{-}$, 
is observed with a statistical significance of more than $30\sigma$, and the decays 
$K^{*}(1680)^{-}\to\Km\eta$ and $K^{*}_2(1980)^{-}\to\Km\eta$ are observed for the first time 
and have statistical significances of $16\sigma$ and $17\sigma$, respectively. 

We extract the signal yield from the $D^0$ invariant mass distribution in $1.78~{\rm GeV}/c^2<M<1.94~{\rm GeV}/c^2$ and $|Q-5.85|<1.0$ MeV/$c^2$, and obtain for the first time the branching ratio $\frac{\mathcal{B}(D^0\to\Km\pip\eta)}{\mathcal{B}(D^0\to\Km\pip)}=0.500\pm0.002{\rm(stat)}\pm0.020{\rm(syst)}\pm0.003{\rm (\mathcal{B}_{PDG})}$, which corresponds to $\mathcal{B}(D^0\to\Km\pip\eta)=(1.973\pm0.009{\rm(stat)}\pm0.079{\rm(syst)}\pm0.018{\rm (\mathcal{B}_{PDG})})\%$. 
Then utilizing the world average branching fractions of intermediate resonant decays, the relative branching ratio $\frac{\mathcal{B}(K^*(1680)^-\to\Km\eta)}{\mathcal{B}(K^*(1680)^-\to\Km\piz)}$ is determined to be $0.11\pm0.02{\rm(stat)}^{+0.06}_{-0.04}{\rm(syst)}\pm0.04{\rm(\mathcal{B}_{\text{PDG}})}$. 
This is not consistent with the theoretical prediction under an assumption of a pure $1^{3}D_1$ 
state~\cite{bib:K2st1980}. We also determine the product of branching fraction 
$\mathcal{B}(D^0\to[K_2^{*}(1980)^-\to\Km\eta]\pip)=(2.2^{+1.7}_{-1.9})\times10^{-4}$. 
For $a_0(980)^+$, we confirm the $\pi\eta^{\prime}$ contribution in the three-channel Flatt\'{e} model 
with a statistical significance of $10.1\sigma$. We have also determined the branching fraction 
$\mathcal{B}(D^0\to\bar{K}^{*}(892)^0\eta)=(1.41^{+0.13}_{-0.12})\%$, which is consistent with, 
and more precise than, the current world average of $(1.02\pm0.30)\%$. 
It deviates from the various theoretical predictions of (0.51-0.92)\%~\cite{bib:thoery} with a significance of more than $3\sigma$.

\section{Measurement of Branching Fractions of $\Lambda_c^+\to\eta\Lambda\pip$, $\eta\Sigma^0\pip$, $\Lambda(1670)\pip$, and $\eta\Sigma(1385)^+$}
\begin{figure}[!htpb]
  \begin{centering}
  \begin{overpic}[width=0.45\textwidth,height=0.35\textwidth]{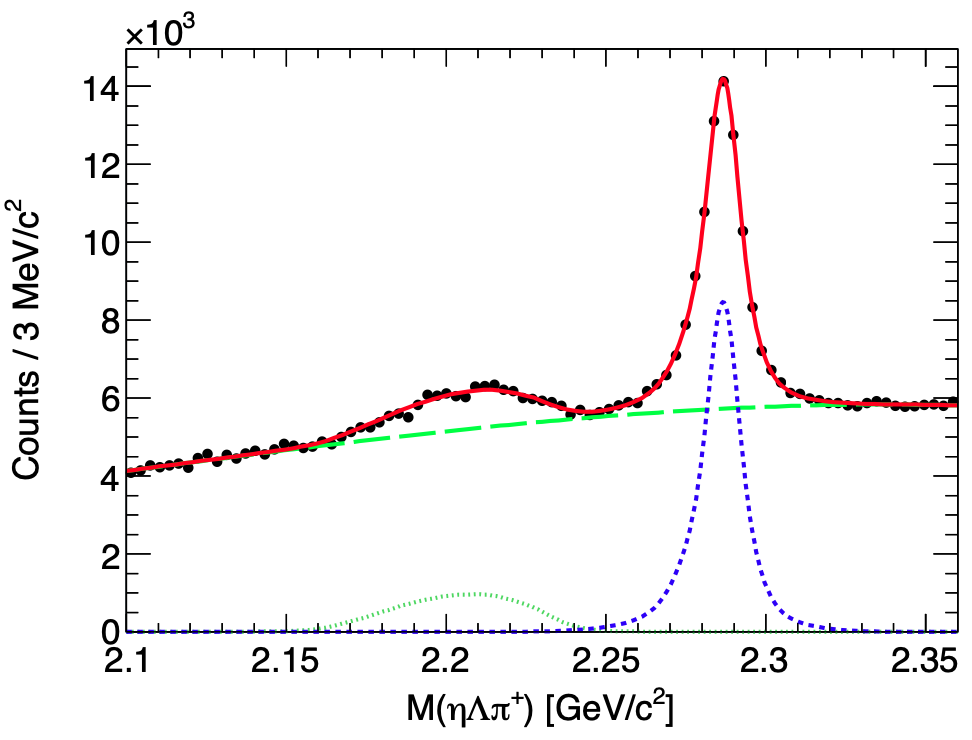}
  \put(20, 60){\large(a)}
  \end{overpic}~~~~%
  \begin{overpic}[width=0.5\textwidth,height=0.34\textwidth]{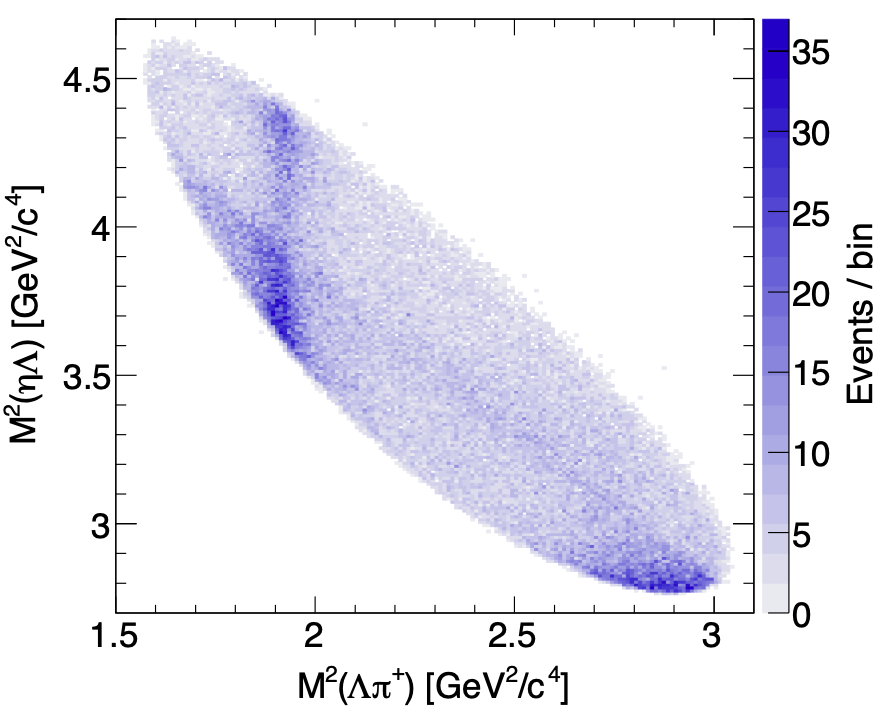}
  \put(70, 52){\large(b)}
  \end{overpic}\\
  \begin{overpic}[width=0.45\textwidth,height=0.35\textwidth]{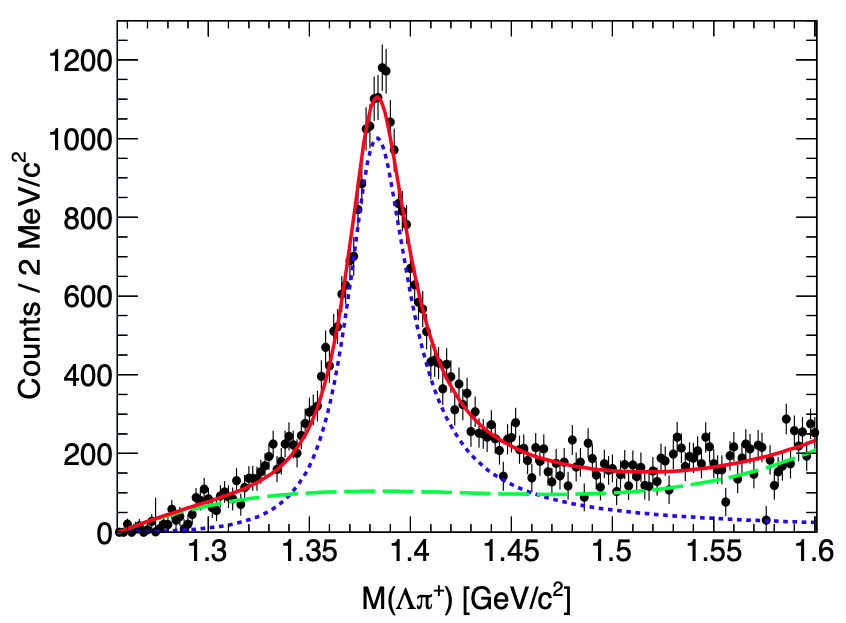}
  \put(20, 60){\large(c)}
  \end{overpic}~~~~%
  \begin{overpic}[width=0.46\textwidth,height=0.355\textwidth]{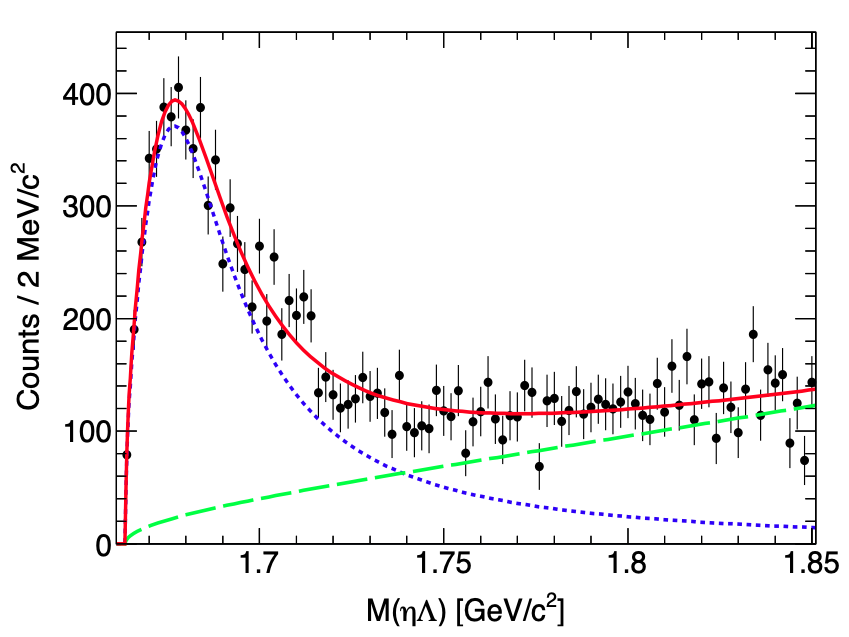}
  \put(70, 60){\large(d)}
  \end{overpic}%
  \vskip-5pt
  \caption{\label{fig:LcToEtaLambPi0} Top figures are (a) the invariant mass of $\eta\Lambda\pip$ and (b) its Dalitz plot in signal region. Bottom figures are fits to the $\Lambda_c^+$ yield in the (c) $M(\eta\Lambda)$ and (d) $M(\Lambda\pip)$ spectra, where the curves indicate the total fit result (solid red), the signal modeled with a relativistic Breit-Wigner function (dashed blue), and the background (long-dashed green).}
  \end{centering}
\end{figure}

The branching fractions of weakly decaying charmed baryons provide a way to study both strong and weak interactions. 
The $\Lambda_c^+\to\eta\Lambda\pip$ decay mode is especially interesting since it has been suggested that it is an ideal decay mode to study the $\Lambda(1670)$ and $a_0(980)$ because, for any combination of two particles in the final state, the isospin is fixed. 
Based on a 980 $\rm fb^{-1}$ data sample, the branching fractions of $\Lambda_c^+\to\eta\Lambda\pip$, $\eta\Sigma^0\pip$, $\Lambda(1670)\pip$, and $\eta\Sigma(1385)^+$ are measured~\cite{bib:arxiv200811575}. The $M(\eta\Lambda\pip)$ spectrum is shown in Fig.~\ref{fig:LcToEtaLambPi0} (a). The $\Lambda_c^+\to\eta\Sigma^0\pip$ is observed indirectly as a feed-down component and it has efficiency-corrected yield $N_{cor}=(3.05\pm0.16)\times10^5$. 
Considering $\Lambda_c^+\to\eta\Lambda\pip$ and $\Lambda_c^+\to p\Km\pip$ have sufficiently large statistic, the yields in individual bins of Dalitz plots are determined: $N_{cor}(\eta\Lambda\pip)=(7.41\pm0.07)\times10^5$ and $N_{cor}(p\Km\pip)=(1.005\pm0.001)\times10^7$. 
Finally, the branching ratios of $\Lambda_c^+\to\eta\Lambda\pip$ and $\Lambda_c^+\to\eta\Sigma^0\pip$ relative to $\Lambda_c^+\to p\Km\pip$ are $0.293\pm0.003\pm0.014$ and $0.120\pm 0.006\pm 0.006$, where the uncertainties are statistical and systematic, respectively. 

On the Dalitz plot of $\Lambda_c^+\to\eta\Lambda\pip$ shown in Fig.~\ref{fig:LcToEtaLambPi0} (b), bands corresponding to $\Lambda_c^+\to\Lambda(1670)\pip/\eta\Sigma(1385)^+$ resonant sub-channels are seen clearly, along with $\Lambda_c^+\to\Lambda a_0(980)^+$. 
For every 2 MeV/$c^2$ bin of $M(\eta\Lambda)$ and $M(\Lambda\pip)$ distributions, the $\Lambda_c^+$ yield is obtained by fitting $M(\eta\Lambda\pip)$. Then, a relativistic Breit-Wigner with momentum-dependent width is used to describe the S-wave $\Lambda(1670)$ and the P-wave $\Sigma(1385)$, as shown in Fig.~\ref{fig:LcToEtaLambPi0} (c, d). Then, we determine the relative branching ratio 
$\frac{\mathcal{B}(\Lambda_c^+\to[\Lambda(1670)\to\eta\Lambda]\pip)}{\mathcal{B}(\Lambda_c^+\to p\Km\pip)}=(5.54\pm0.29\pm0.73)\%$ and $\frac{\mathcal{B}(\Lambda_c^+\to\eta\Sigma(1385)^+)}{\mathcal{B}(\Lambda_c^+\to p\Km\pip)}=0.192\pm 0.006\pm 0.016$. 
Finally after using the world averaged $\mathcal{B}(\Lambda_c^+\to p\Km\pip)$, we have $\mathcal{B}(\Lambda_c^+\to[\Lambda(1670)\to\eta\Lambda]\pip)=(3.48\pm0.19\pm0.46\pm0.18)\times10^{-3}$ and $\mathcal{B}(\Lambda_c^+\to\eta\Sigma(1385)^+)=(1.21\pm0.04\pm0.10\pm0.06)\%$, where the first two uncertainties are statistical and systematic uncertainties, and the third uncertainty is from $\mathcal{B}(\Lambda_c^+\to p\Km\pip)$.

\section{First determination of the Spin and Parity of $\Xi_c(2970)^+$}
The unclear theoretical situation motivates an experimental determination of spin and parity of a charmed-strange baryon $\Xi_c(2970)$, which provides important information to test predictions and help decipher its nature. 
Using a 980 $\rm fb^{-1}$ data sample, the spin and parity of a charmed-strange baryon $\Xi_c(2970)^+$ is measured~\cite{bib:arxiv200714700} by (1) studies of the helicity angle distributions, $\theta_h$ of $\Xi_c(2970)^0$ and $\theta_c$ of $\Xi_c(2645)^0$ in $\Xi_c(2970)^+\to \Xi_c(2645)^0\pip\to \Xi_c^+\pim\pip$, and (2) a measurement of the $\Xi_c(2970)^+$ decay branching ratio $R=\mathcal{B}(\Xi_c(2970)^+\to\Xi_c(2645)^0\pip)/\mathcal{B}(\Xi_c(2970)+\to\Xi_c^{\prime0}\pip)$.

\begin{figure}[!htpb]
  \begin{centering}
  \begin{overpic}[width=0.45\textwidth,height=0.35\textwidth]{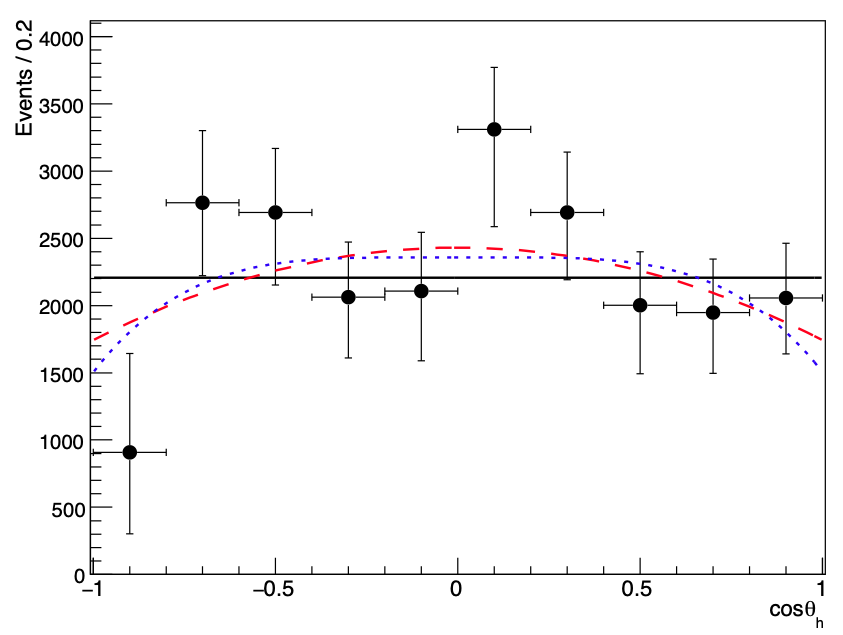}
  \end{overpic}~~~~%
  \begin{overpic}[width=0.45\textwidth,height=0.35\textwidth]{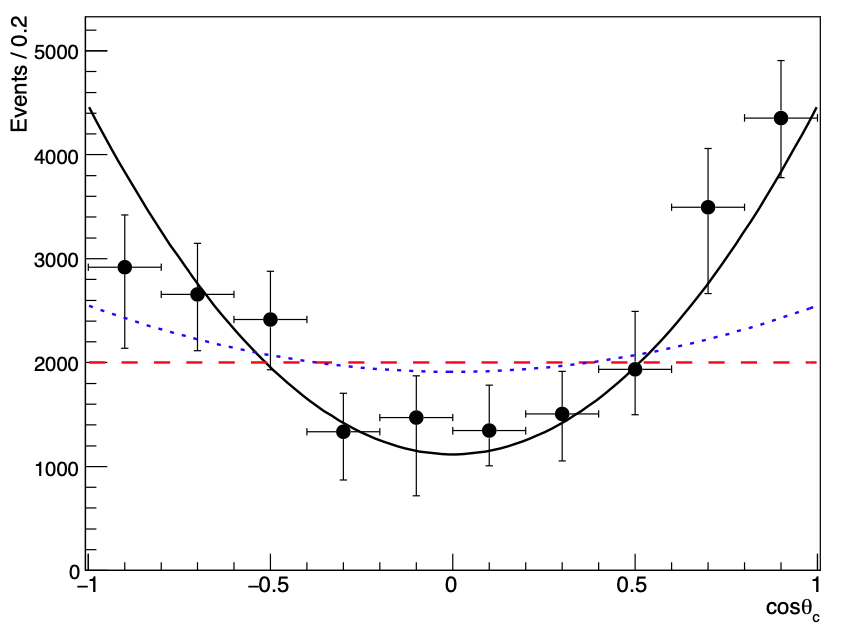}
  \end{overpic}
  \vskip-5pt
  \caption{\label{fig:Xic2645} The yields on the cosine of helicity angle of $\Xi_c(2970)^0$ (left, $J=1/2$ for solid black; $J=3/2$ for dashed red; $J=5/2$ for dotted blue) and on cosine of helicity angle of $\Xi_c(2645)^0$ (right, $J^P=1/2^+$ for solid black; $J=3/2^-$ for dashed red; $J=5/2^+$ for dotted blue) in $\Xi_c(2970)^+\to\Xi_c(2645)^0\pip$ decay.}
  \end{centering}
\end{figure}

The angular distribution are obtained by dividing the data into 10 equal bins for $\cos\theta_h$ and $\cos\theta_c$, each within intervals of 0.2. 
The yield of $\Xi_c(2970)^+\to\Xi_c(2645)^0\pip$ for each bin is obtained by fitting the invariant-mass distribution of $M(\Xi_c^+\pim\pip)$ for the $\Xi_c(2645)^0$ signal region (within 5 MeV/$c^2$ of $\Xi_c(2645)^0$ nominal mass) and sidebands (interval from 15 to 25 MeV/$c^2$ away from $\Xi_c(2645)^0$ nominal mass). 
The background-subtracted and efficiency-corrected yield distribution in Fig.~\ref{fig:Xic2645} is fitted with expected decay-angle distributions $W_J$ for different spin hypotheses. The best fit is for $J=1/2$, while others are excluded with small significance, which shows inconclusive result. For helicity angle $\theta_c$, with an assumption that the lowest partial wave dominates, the expected angular correlation $W(\theta_c)$ is used to describe the distribution. Finally the $J^P=1/2^+$ hypothesis is better than $3/2^-$ or $5/2^+$ ones at the level of $5.1\sigma$ or $4.0\sigma$.

\begin{figure}[!htpb]
  \begin{centering}
  \begin{overpic}[width=0.45\textwidth,height=0.35\textwidth]{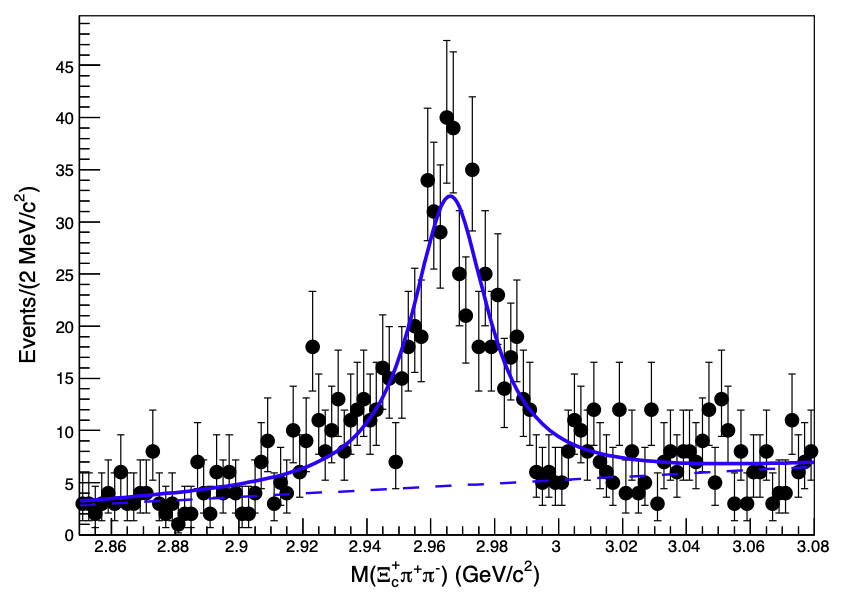}
  \end{overpic}~~~~%
  \begin{overpic}[width=0.46\textwidth,height=0.35\textwidth]{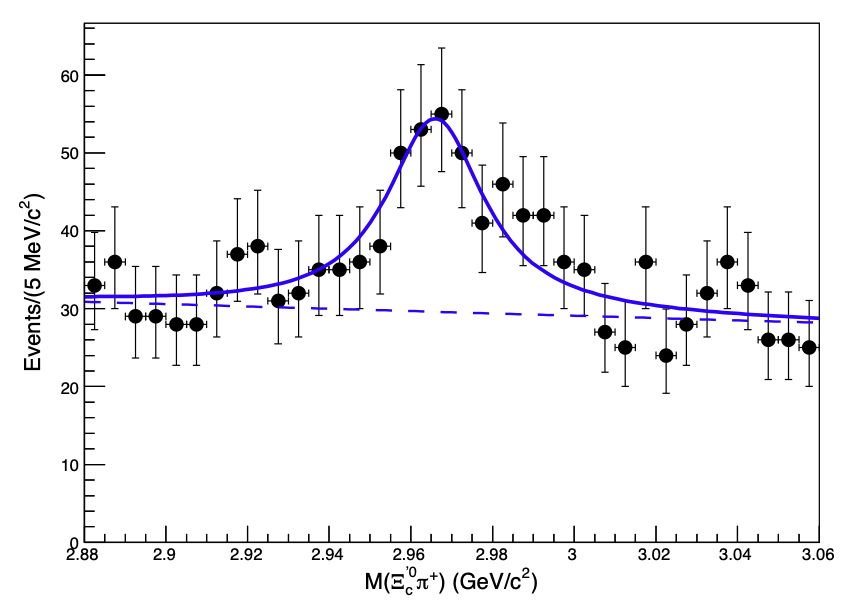}
  \end{overpic}%
  \vskip-5pt
  \caption{\label{fig:Xic2645_2} $\Xi_c^+\pim\pip$ invariant-mass distribution for $\Xi_c(2970)^+\to\Xi_c(2645)^0\pip\to\Xi_c^+\pim\pip$, and $\Xi_c^{\prime0}\pip$ invariant-mass distribution for $\Xi_c(2970)^+\to\Xi_c^{\prime0}\pip\to\Xi_c^0\gamma\pip$. The fit result (solid blue curve) is presented along with the background (dashed blue curve)}
  \end{centering}
\end{figure}

The parity of $\Xi_c(2970)^+$ is established~\cite{bib:arxiv200714700} from the ratio between $\mathcal{B}(\Xi_c(2970)^+\to\Xi_c(2645)^0\pip)$ and $\mathcal{B}(\Xi_c(2970)^+\to\Xi_c^{\prime0}\pip)$ by $R=\frac{N^{*}}{\epsilon^{*}N(\Xi_c^+)/\epsilon^+}/\frac{N^{\prime}}{\sum_i\epsilon_i^{\prime}N(\Xi_c^0)_i/\epsilon_i^0}$, where $\Xi_{c}^0$ uses two modes, $\Xi^{-}\pip$ and $\Omega^-K^+$. The yields $N^{*,\prime}$ are obtained by fitting the invariant-mass distributions in Fig.~\ref{fig:Xic2645_2}. 
Finally we have $R=1.67\pm0.29(stat)^{+0.15}_{-0.09}(syst)\pm0.25(IS)$, where the last uncertainty is due to possible isospin-symmetry-breaking effects (15\%). Heavy-quark spin symmetry (HQSS) predicts $R=1.06$ (0.26) for a $1/2^+$ state with the spin of the light-quark degrees of freedom $s_l=0$ (1)~\cite{bib:PRD75d014006}.
Our result favors a positive-parity assignment with $s_l=0$. We note that HQSS predictions could be larger than the quoted value by a factor of $\sim2$ with higher-order terms in ($1/m_c$)~\cite{bib:PRD53d231}, so our result is consistent with the HQSS prediction for $J^P(s_l)=1/2^+(0)$. 

\section{Summary}
Belle experiment has achieved the fruitful productions of flavor physics to date. Some selected recent charm results are presented, including charm mixing parameter $y_{CP}$ in $CP$-odd decay $D^0\to\Ks\omega$, hadronic decays $D^0\to\Km\pip\eta$ and $\Lambda_c^+\to\eta\Lambda\pip/\eta\Sigma^0\pip$, first determination of the spin and parity of $\Xi_c(2970)^+$.
More charming charm results from Belle will come out in near future. As a summary, I would like to say, "Belle is not only keeping alive but still keeping energetic, together with its upgraded experiment Belle II who is under a rapid growth."

\end{document}